\documentclass[twocolumn,american,aps, prl, superscriptaddress, showpacs]{revtex4-1}
\usepackage[T1]{fontenc}
\usepackage[latin9]{inputenc}
\usepackage{float}
\usepackage{booktabs}
\usepackage{textcomp}
\usepackage{graphicx}
\usepackage{color}

\makeatletter

\providecommand{\tabularnewline}{\\}

\@ifundefined{textcolor}{}
{%
 \definecolor{BLACK}{gray}{0}
 \definecolor{WHITE}{gray}{1}
 \definecolor{RED}{rgb}{1,0,0}
 \definecolor{GREEN}{rgb}{0,1,0}
 \definecolor{BLUE}{rgb}{0,0,1}
 \definecolor{CYAN}{cmyk}{1,0,0,0}
 \definecolor{MAGENTA}{cmyk}{0,1,0,0}
 \definecolor{YELLOW}{cmyk}{0,0,1,0}
 }

\makeatother

\usepackage{babel}

\begin{document}

\title{Observation of isotonic symmetry for enhanced quadrupole collectivity\\
in neutron-rich $^{62,64,66}$Fe isotopes at $N$~=~40}

\author{W.~Rother}
\affiliation{Institut f\"ur Kernphysik der Universit\"at zu K\"oln, D-50937 K\"oln, Germany}

\author{A.~Dewald}
\affiliation{Institut f\"ur Kernphysik der Universit\"at zu K\"oln, D-50937 K\"oln, Germany}

\author{H.~Iwasaki}
\affiliation{National Superconducting Cyclotron Laboratory, Michigan State University,
East Lansing, Michigan 48824, USA}
\affiliation{Department of Physics and Astronomy, Michigan State University, East
Lansing, Michigan 48824, USA}

\author{S.~M.~Lenzi}
\affiliation{Dipartimento di Fisica dell\textquoteright{} Universit\'a and INFN,
Sezione di Padova, Padova, Italy}

\author{K.~Starosta}
\affiliation{Department of Chemistry, Simon Fraser University, Burnaby BC V5A 1S6, Canada}

\author{D.~Bazin}
\affiliation{National Superconducting Cyclotron Laboratory, Michigan State University,
East Lansing, Michigan 48824, USA}

\author{T.~Baugher}
\affiliation{National Superconducting Cyclotron Laboratory, Michigan State University,
East Lansing, Michigan 48824, USA}
\affiliation{Department of Physics and Astronomy, Michigan State University, East
Lansing, Michigan 48824, USA}

\author{B.~A.~Brown}
\affiliation{National Superconducting Cyclotron Laboratory, Michigan State University,
East Lansing, Michigan 48824, USA}
\affiliation{Department of Physics and Astronomy, Michigan State University, East
Lansing, Michigan 48824, USA}

\author{H.~L.~Crawford}
\affiliation{National Superconducting Cyclotron Laboratory, Michigan State University,
East Lansing, Michigan 48824, USA}
\affiliation{Department of Chemistry, Michigan State University, East
Lansing, Michigan 48824, USA}

\author{C.~Fransen}
\affiliation{Institut f\"ur Kernphysik der Universit\"at zu K\"oln, D-50937 K\"oln, Germany}


\author{A.~Gade}
\affiliation{National Superconducting Cyclotron Laboratory, Michigan State University,
East Lansing, Michigan 48824, USA}
\affiliation{Department of Physics and Astronomy, Michigan State University, East
Lansing, Michigan 48824, USA}

\author{T.~N.~Ginter}
\affiliation{National Superconducting Cyclotron Laboratory, Michigan State University,
East Lansing, Michigan 48824, USA}

\author{T.~Glasmacher}
\affiliation{National Superconducting Cyclotron Laboratory, Michigan State University,
East Lansing, Michigan 48824, USA}
\affiliation{Department of Physics and Astronomy, Michigan State University, East
Lansing, Michigan 48824, USA}

\author{G.~F.~Grinyer}
\affiliation{National Superconducting Cyclotron Laboratory, Michigan State University,
East Lansing, Michigan 48824, USA}

\author{M.~Hackstein}
\affiliation{Institut f\"ur Kernphysik der Universit\"at zu K\"oln, D-50937 K\"oln, Germany}

\author{G.~Ilie}
\affiliation{Wright Nuclear Structure Laboratory, Yale University, New Haven,
Connecticut 06520, USA}
\affiliation{National Institute of Physics and Nuclear Engineering, 76900 Bucharest,
Romania}

\author{J.~Jolie}
\affiliation{Institut f\"ur Kernphysik der Universit\"at zu K\"oln, D-50937 K\"oln, Germany}

\author{S.~McDaniel}
\affiliation{National Superconducting Cyclotron Laboratory, Michigan State University,
East Lansing, Michigan 48824, USA}
\affiliation{Department of Physics and Astronomy, Michigan State University, East
Lansing, Michigan 48824, USA}

\author{D.~Miller}
\affiliation{Department of Physics and Astronomy, University of Tennessee, Knoxville, TN 37996, USA}

\author{P.~Petkov}
\affiliation{Institut f\"ur Kernphysik der Universit\"at zu K\"oln, D-50937 K\"oln, Germany}
\affiliation{Institute for Nuclear Research and Nuclear Energy, BAS, 1784 Sofia, Bulgaria}

\author{Th.~Pissulla}
\affiliation{Institut f\"ur Kernphysik der Universit\"at zu K\"oln, D-50937 K\"oln, Germany}

\author{A.~Ratkiewicz}
\affiliation{National Superconducting Cyclotron Laboratory, Michigan State University,
East Lansing, Michigan 48824, USA}
\affiliation{Department of Physics and Astronomy, Michigan State University, East
Lansing, Michigan 48824, USA}

\author{C.~A.~Ur}
\affiliation{Dipartimento di Fisica dell\textquoteright{} Universit\'a and INFN,
Sezione di Padova, Padova, Italy}

\author{P.~Voss}
\affiliation{National Superconducting Cyclotron Laboratory, Michigan State University,
East Lansing, Michigan 48824, USA}
\affiliation{Department of Physics and Astronomy, Michigan State University, East
Lansing, Michigan 48824, USA}

\author{K.~A.~Walsh}
\affiliation{National Superconducting Cyclotron Laboratory, Michigan State University,
East Lansing, Michigan 48824, USA}
\affiliation{Department of Physics and Astronomy, Michigan State University, East
Lansing, Michigan 48824, USA}

\author{D.~Weisshaar}
\affiliation{National Superconducting Cyclotron Laboratory, Michigan State University,
East Lansing, Michigan 48824, USA}

\author{K.-O.~Zell }
\affiliation{Institut f\"ur Kernphysik der Universit\"at zu K\"oln, D-50937 K\"oln, Germany}

\begin{abstract}
The transition rates for the $2_{1}^{+}$ states 
in $^{62,64,66}$Fe 
were studied using the Recoil Distance 
Doppler-Shift technique applied to projectile Coulomb excitation reactions.
The deduced E2 strengths illustrate the enhanced collectivity of the
neutron-rich Fe isotopes up to $N$~=~40.
The results are interpreted by the generalized concept of valence proton
symmetry 
which describes the
evolution of nuclear structure
around $N$~=~40 as
 governed by the number of valence protons 
with respect to 
$Z$~$\approx$~30.
The 
deformation suggested by the experimental data
is reproduced by state-of-the-art shell calculations with a new
effective interaction developed for the $fpgd$ valence space.
\end{abstract}


\maketitle


Configuration mixing is one of the most important concepts of 
quantum mechanics.
Orbital mixing of electron configurations in organic molecules
provides extra stability to the system; 
quark mixing accounts for rare meson decays violating the $CP$ symmetry.

Configuration mixing in nuclear states provides a rich aspect 
of atomic nuclei, since it can considerably enhance amplitudes 
of collective excitation modes.
Quadrupole deformation in nuclei is one such example and is known to evolve dramatically. Pronounced collectivity or deformation is normally observed in nuclei away from closed shells. It can be correlated with the dimension of the valence space, which is the product of valence proton and neutron numbers. 
This highlights a unique feature of atomic nuclei as a two fermionic quantum system. 

Recently, accumulated evidence has shown that magic numbers of nuclei can change far from the $\beta$ stability line~\cite{Otsu10,Cau02}.
This has challenged our familiar picture of the shell closure,
motivating the quest to understand the evolution of collectivity 
and mixing properties of associated configurations at extreme isospin.


In the present work, we measured the mean lifetime~($\tau$) of 
the first 2$^{+}$ (2$^{+}_{1}$) 
states of the neutron-rich $^{62,64,66}$Fe isotopes
at and around the neutron number $N$~=~40.
As a direct measure of quadrupole collectivity,
we deduced the reduced E2 transition probabilities $B$(E2;2$^{+}$$\to$0$^{+}$),
which are inversely proportional to $\tau$.
Fe isotopes, spanning from the most neutron-deficient 
two-proton emitter $^{45}$Fe$_{19}$~\cite{Giov02}, 
via the most abundant isotope $^{56}$Fe$_{30}$, 
to the most neutron-rich $^{74}$Fe$_{48}$ hitherto observed~\cite{Ohnishi10}, 
provide an intriguing testing ground for theoretical models to
comprehend the development and evolution of quadrupole collectivity over a wide range of neutron numbers.
The $B$(E2) values should follow an inverted parabola
when plotted as a function of valence neutrons going from the closed
shell at $N$~=~20 towards the next closed shell at $N$~=~28 and then again
from $N$~=~28 to $N$~=~50. The magicity at $N$~=~50 is 
still an open question in the regime of large neutron excess.
%
A key issue is the degree of quadrupole 
collectivity manifested in $^{66}$Fe$_{40}$
at the $N$~=~40 harmonic-oscillator shell closure.
If the $N$~=~40 shell closure is absent 
$-$ as suggested by the low excitation energy $E$(2$^{+}$) 
of the 2$^{+}_{1}$ state in $^{66}$Fe~\cite{Han99} $-$ 
one can expect the maximum collectivity at $N$~$\approx$~40, 
which corresponds to the middle of the magic numbers 28 and 50.
In fact, a recent study indicated a sudden increase of $B$(E2) from 
$^{62}$Fe$_{36}$ to $^{64}$Fe$_{38}$~\cite{Lju10}.
However, this picture is in contrast to a 
doubly magic character of the neighboring $N$~=~40
isotone $^{68}$Ni$_{40}$,
where a steep decrease of $B$(E2) 
has been observed from $^{66}$Ni to $^{68}$Ni~\cite{Sor02}.
%
%
%
%
%
%
%

%
In this Letter,
we investigate the interplay between 
quadrupole collectivity in $^{62,64,66}$Fe 
and the configuration mixing of valence nucleons
in terms of the semi-empirical scheme of the valence proton 
symmetry~\cite{Casten93,Dewald2008} 
as well as state-of-the-art shell model calculations.
Based on new $^{66}$Fe data and additional ones that are more precise than those
given in~\cite{Lju10}, we provide a new insight into
the complex evolution of
collectivity in the vicinity of $N$~=~40.



The experiment was performed at National Superconducting Cyclotron
Laboratory (NSCL), Michigan State University.
The lifetimes of the 2$^{+}_{1}$ states in $^{62,64,66}$Fe 
were measured by means of the Recoil Distance Doppler Shift (RDDS)
technique applied to intermediate-energy Coulomb excitation.
%
%
%
This technique has been demonstrated for medium-heavy nuclei with 
$Z$~$\approx$~50~\cite{Chester2006,Dewald2008}. 
In this work, we extended the RDDS measurements 
to lighter Fe projectiles with $Z$~=~26 
by optimizing the target-degrader combination 
of the K\"oln/NSCL plunger device~\cite{Chester2006,Dewald2008,Starosta2007}.
By making use of intense secondary beams provided by the Coupled Cyclotron Facility 
at NSCL, the Segmented Germanium Array (SeGA) for gamma-ray detection, and
the S800 spectrograph~\cite{Bazin03} 
for identification of reaction products,
we realized the lifetime measurements of neutron-rich Fe isotopes up to $N$~=~40.

Secondary beams of Fe were produced by fragmentation of
a primary $^{76}$Ge beam at 130 $A$MeV incident on $^{9}$Be targets.
The A1900 fragment separator~\cite{Morrissey2003}
was used to purify the fragments.
The resulting beam was 
$\approx$85$\%$ $^{62}$Fe (at a typical rate of 3.6$\times$10$^{4}$~pps
and an incident energy of 97.8~$A$MeV), 
$\approx$65$\%$ $^{64}$Fe (6$\times$10$^{3}$~pps, 95.0~$A$MeV), and 
$\approx$25$\%$ $^{66}$Fe (1$\times$10$^{3}$~pps, 88.3~$A$MeV) for
each setting. 

The beams were directed onto the plunger device, which was mounted
at the target position of the S800 spectrograph. 
The plunger target was a 300~$\mu$m-thick Au foil,
while a 300~$\mu$m~(400~$\mu$m)-thick Nb foil was used as a degrader
for the $^{62,66}$Fe~($^{64}$Fe) measurement.
The beam velocities before and after the degrader were estimated to be
$v$/$c$~=~0.368, 0.322 for $^{62}$Fe,
$v$/$c$~=~0.364, 0.298 for $^{64}$Fe, and
$v$/$c$~=~0.346, 0.291 for $^{66}$Fe, respectively,
 with typical velocity widths (in r.m.s.) of $\approx$1$\%$.
Mass and charge of the incoming beams were identified event-by-event
using the time-of-flight measured as the timing difference between two plastic
scintillators placed in the extended focal plane of the A1900 and 
in the object plane of the S800 analysis beam line. 
Scattered particles were identified by the 
detection system
of the S800~\cite{Bazin03}.

Doppler-shifted $\gamma$ rays were detected by SeGA coupled with
the digital data acquisition system~\cite{DDAS2009}. 
Two rings of 7 and 8 detectors were mounted at forward and backward angles 
of 30$^{\circ}$ and 140$^{\circ}$ relative to the beam axis.
Data were taken at 5 $-$ 7 different target-degrader distances 
over the range from 0 to 20~mm for different Fe isotopes.
The data taken at the largest distances were used to examine the contributions
from Coulomb excitation reactions in the degrader~\cite{Starosta2007,Dewald2008}.



\begin{figure}[tb]
\noindent 
\centering{}
\includegraphics[width=8.5cm]{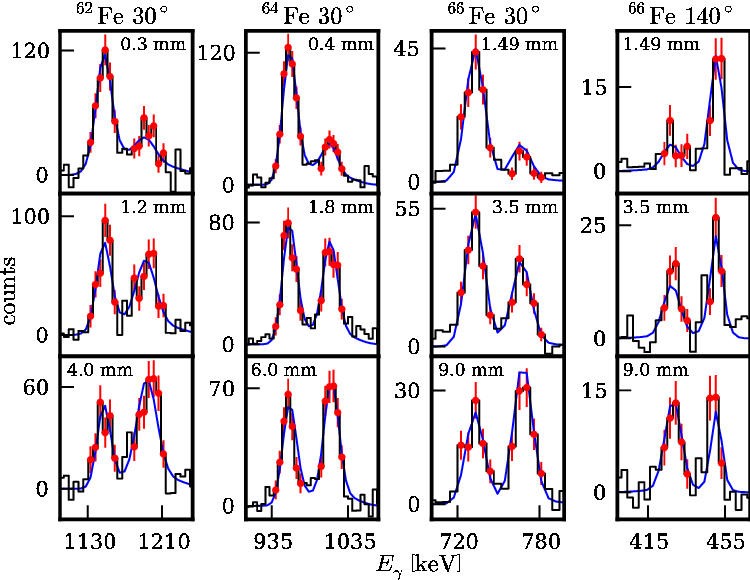}
\caption{
(color online)
Typical results from the $\gamma$-ray line-shape fits 
to the data for the 2$^{+}$$\rightarrow$0$^{+}$ transitions in $^{62,64,66}$Fe.
}
\label{fig:lineshapes}
\end{figure}

Lifetimes were determined by line-shape analysis
in a similar procedure to those developed 
in Refs.~\cite{Starosta2007,Dewald2008}. 
The best-fit results for each Fe isotope
are shown in Fig.~\ref{fig:lineshapes}.
The calculation takes into account 
the velocities of the incoming and outgoing beams,
the energy losses through the target and degrader, and
the detector geometry including effects due to the Lorentz boost.
The energy and angular straggling of the reaction products 
and the energy resolution of the $\gamma$-ray detectors 
are described as a single width parameter of a Gaussian distribution
incorporated in the lineshapes. 
We used four different width parameters for decays before and
after the degrader in the spectra obtained with the forward and
backward rings.
The fits to the data were performed using variable parameters
which include the lifetime of the state of interest, 
the width parameters, a normalization factor, as well as 
the yield ratio between target and degrader. 
The large distance data were found to be consistent with degrader-to-target yield ratios of around 30--40~$\%$ for 
$^{62,64,66}$Fe.

In the present analysis, special care was taken to select sensitive
regions of the spectra which ensure the reliability of our $\chi^{2}$ 
analysis in determining the lifetime. To this end, after fixing the parameters
that characterize the lineshape, we compared the simulated spectra 
to the measured ones  by varying only the lifetimes.
It became obvious that the most sensitive region 
corresponds to the 
$\gamma$-ray energy range closest to the two peak centroids.
It is worth noting that the peak amplitudes are most sensitive to the 
lifetime, whereas the $\chi^2$ value can also be affected by details
of the lineshape.
Therefore, we restricted the $\chi^{2}$ calculation to the optimum regions
as indicated by the data with the error bars in Fig.~\ref{fig:lineshapes}.
We found a higher sensitivity to the lifetime for the spectra obtained 
from the forward-ring detectors. 
We thus first made a fit only to the forward-ring spectra 
to search for the $\chi^2$ minimum over a wide range of the lifetime. 
Later, we determined the lifetime 
from an overall fit around the obtained minimum using the data
taken with both rings. 
The results of $\tau$
are summarized in Table~\ref{Table1}, together with the 
corresponding $B$(E2;2$^{+}$$\rightarrow$0$^{+}$) values.
The final lifetime results are 
consistent with the first results obtained with the forward-ring data.
For $^{62}$Fe, we used only forward-ring data due to a contaminant
in the backward-ring spectra.



\begin{table}[bt]
%
%
\caption{ Comparison between the experimental~(Exp) and theoretical~(SM) 
$E$(2$^{+}$) and $B$(E2) values for $^{62,64,66}$Fe. The lifetimes
($\tau_{\rm Exp}$) are the present results, while the $E$(2$^{+}$)
values are taken from Ref.~\cite{Lun07}. Previous 
$B$(E2) data (Pre)~\cite{Lju10} are also compared.}

\begin{centering}
\begin{tabular}{crccccc}
\hline 
$A$  & $\tau_{\rm Exp}$ [ps]  & \multicolumn{2}{c}{$E$(2$^{+}$) [keV]} & \multicolumn{3}{c}{$B$(E2) [e$^{2}$fm$^{4}$]}
\tabularnewline
 &  & $_{\rm Exp}$  & $_{\rm SM}$  & $_{\rm Exp}$  & $_{\rm Pre}$  & $_{\rm SM}$ \tabularnewline
\hline 
$^{62}$Fe  & 8.0(10)  & 877  & 835  & 198(25)  & 214(26)  & 270 \tabularnewline
\hline 
$^{64}$Fe  & 10.3(10)  & 746  & 747  & 344(33)  & 470$_{-110}^{+210}$  & 344 \tabularnewline
\hline 
$^{66}$Fe  & 39.4(40)  & 575  & 570  &  332(34)  & & 421 \tabularnewline
\hline
\end{tabular}
\par\end{centering}

\label{Table1} 
\end{table}

\begin{figure}[tb]
\noindent 
\centering{}
\includegraphics[width=8cm]{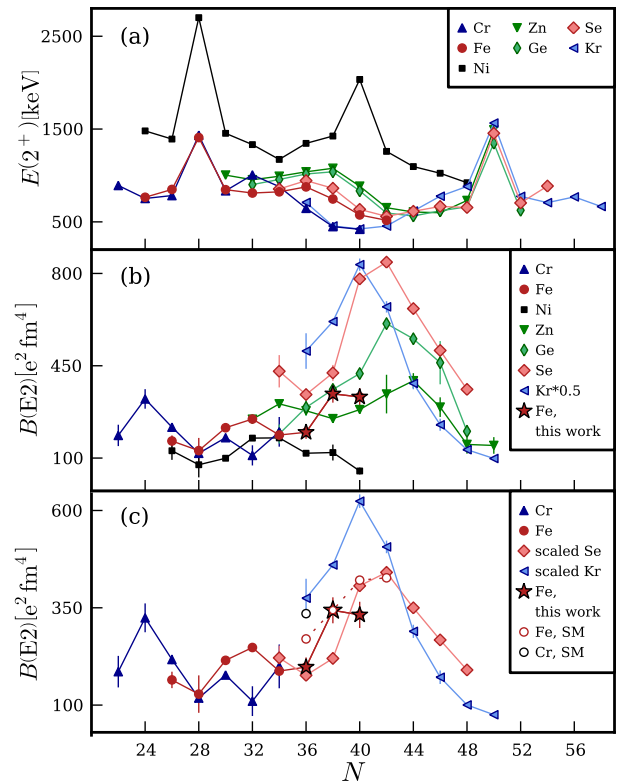}
\caption{
(color online) (a)
Energies $E$(2$^{+}$) and 
(b)~$B$(E2;2$^{+}$$\rightarrow$0$^{+}$) values
for even-even nuclei with $Z$~=~24~--~36 and $N$~=~24~--~56.
The stars indicate the present data of $^{62,64,66}$Fe.
The $B$(E2) data of Kr in (b) are divided by 2.
In (c), the $B$(E2) data of the Fe and Cr isotopes are compared to
the scaled $B$(E2) values for the Se and Kr isotopes, respectively,
as well as the shell model predictions (SM) 
for $^{62,64,66,68}$Fe and $^{60}$Cr isotopes.
}
\label{energies}
\end{figure}

The evolution of collectivity for even-even nuclei with $Z$~=~24~--~36
and $N$~=~22~--~58 is studied in Fig.~\ref{energies}
from the systematic behavior
of the energy 
of the 2$^{+}_{1}$ states (Fig.~\ref{energies} (a))
and the $B$(E2;2$^{+}$$\to$0$^{+}$) values 
(Fig.~\ref{energies} (b))~\cite{ENSDF,Hotel06,Aoi09,Gad10,
Pad05,Bue05,Adr09,Rza00,Jon06,Obe09,Lju08,Wal09,Lun07}.
Only Ni isotopes reveal characteristics of shell closure 
at $N$~=~28, 40 as inferred from large $E$(2$^{+}$) 
and small $B$(E2) values.
The $N$~=~40 shell closure vanishes 
for all the other even-even isotopes, 
where an enhanced collectivity is evident.  
%
The present $B$(E2) data confirm the onset of collectivity
for $^{62}$Fe and $^{64}$Fe recently reported~\cite{Lju10} (see Table~\ref{Table1}) 
and represent the first evidence that the enhanced
collectivity persists in $^{66}$Fe at $N$~=~40. 
This tendency
is consistent with the other neighboring
isotopes with $Z$~$\neq$~28. 
A large deformation parameter 
of $\beta_2$~$\approx$~0.28 is calculated from
the measured $B$(E2) of $^{64,66}$Fe.


Interestingly, one can see in Fig.~\ref{energies}~(a) notable similarities in the $E$(2$^{+}$) behavior 
for the pairs of Fe~($Z$~=~26)/Se~($Z$~=~34),
and Zn~($Z$~=~30)/Ge~($Z$~=~32)~\cite{Hotel06}. 
An incipient similarity in $E$(2$^{+}$) is also suggested for the Cr~($Z$~=~24)/Kr~($Z$~=~36) pair.  
The enhancement of collectivity is also noticeable
from the decrease of $E$(2$^{+}$) as one moves away from the proton number $Z$~$\approx$~30. 
It starts at $N$~=~36 for Cr/Kr, $N$~=~38 for Fe/Se 
and $N$~=~40 for Zn/Ge, respectively. These striking similarities suggest an isotonic symmetry in the development of collectivity for each of these pairs.

To further investigate the underlying symmetry inferred from
the $E$(2$^{+}$) behavior, we 
turn to 
the valence proton symmetry 
(VPS)~\cite{Casten93,Dewald2008}.
The VPS scheme correlates a pair of isotones
that have the same number of valence proton holes and particles
with respect to closed shells.
The same valence space for correlated nuclei results in
a similar degree of collectivity and its evolution,
as shown, for example, in the region of 50~$<$~$N$~$<$~82 and
$Z$~$\approx$~50~\cite{Dewald2008}.
In this work, we generalized the VPS concept since the symmetry 
appears with respect to the proton number $Z$~=~30 instead of
the canonical magic number $Z$~=~28.
Here, we consider proton holes in the 1$f_{7/2}$ orbital below $Z$~=~28 
and protons in the 1$f_{5/2}$ orbital above the $Z$~=~32 subshell closure.
This is justified by the fact that the 2$p_{3/2}$ orbital is isolated
%
%
%
%
%
%
%
%
between the 1$f_{7/2}$ and 1$f_{5/2}$ orbitals. 
The 2$p_{3/2}$ orbital, if it is fully occupied, has
no impact on the evolution of collectivity,
and hence, for the VPS scheme, the effect of a fully empty or occupied
2$p_{3/2}$ orbital is identical.
Consequently, Se~(Kr) can be a valence partner of Fe~(Cr).
%
%
%
%

To test the modified scheme of the VPS,
in Fig.~\ref{energies} (c) we compare the
existing $B$(E2) data for the Fe/Se and Cr/Kr pairs, including
the present data of $^{62,64,66}$Fe. 
The charge and mass differences are considered in the comparison 
by scaling $B$(E2) data for the heavier isotopes (Se and Kr) 
with the factor of $S$~=~($Z_{L}$/$Z_{H}$)$^{2}$$\times$($A_{L}$/$A_{H}$),
where $Z_{H}$ and $A_{H}$ ($Z_{L}$ and $A_{L}$)
denote the atomic and mass numbers of
heavier (lighter) pair partners, respectively~\cite{Dewald2008}.
While the data are still scarce for Cr/Kr, the present 
data of $^{62,64,66}$Fe clearly agree in magnitude with the overall trend 
of the Se isotopes, proposing the validity of the generalized VPS scheme.
Consequently, one can expect large collectivity for further neutron-rich
Fe and Cr isotopes beyond $N$~$=$~40, as illustrated by the 
scaled $B$(E2) data for the Se and Kr isotopes.




The symmetric evolution of collectivity above and below the proton
number $Z$~=~30 can be understood 
in a shell-model picture~\cite{Lun07},
in terms of an energy decrease of the neutron shell gap at $N$~=~40 
caused by the proton-neutron monopole tensor interaction.
Nuclei above $Z$~=~30 
become more deformed when approaching $N$~=~$Z$~=~40 
$^{80}$Zr. 
Filling protons in the 1$f_{5/2}$ orbital
lowers the neutron $1g_{9/2}$ and $2d_{5/2}$ orbitals.
$Vice$ $versa$, for the light isotones with $Z$~$\leq$~26 that have
proton holes in the 1$f_{7/2}$ orbital, the attractive 
interaction with the neutron 1$f_{5/2}$ orbital and the 
repulsive interaction with the $1g_{9/2}$ and $2d_{5/2}$ orbitals
get weaker. 
In both cases, the shell gap at $N$~=~40 reduces consistently,
enhancing quadrupole collectivity.
A similar effect is produced on the $Z$~=~28 gap.
This is consistent with the ground-state spin parities 
of the $N$~=~39 isotones~\cite{ENSDF}
( $^{77}$Sr~:~5/2$^{+}$, $^{75}$Kr~:~5/2$^{+}$, $^{73}$Se~:~9/2$^{+}$,
$^{71}$Ge~:~1/2$^{-}$, $^{69}$Zn~:~1/2$^{-}$, $^{67}$Ni~:~(1/2)$^{-}$ ),
where the intruder $1g_{9/2}$ (or $2d_{5/2}$) contribution is evident 
only for the heavier isotones away from $Z$~=~30 that have 
positive-parity ground states.

To allow a quantitative discussion of the intruder neutron configurations
for the lighter valence partners, we have performed shell model 
calculations~\cite{LNPS}
and compare them with the present $B$(E2) data for $^{62,64,66}$Fe.
The natural choice for the 
model space is the 
$fp$ shell 
(20~$\leq$~$Z$~$\leq$~40) for protons and 
the $ 2p_{3/2}1f_{5/2} 2p_{1/2} 1g_{9/2}$ space for neutrons 
(28~$\leq$~$N$~$\leq$~50).
However, shell model calculations in this model space
fail to fit the $^{66}$Fe 2$^+_{1}$ state~\cite{Lun07}, while they can reproduce the level schemes of $^{62-64}$Fe rather well.
To reproduce the quadrupole collectivity in this mass region, the inclusion of the neutron $2d_{5/2}$ orbital is necessary~\cite{Cau02, LNPS}.
Recently, a new effective interaction for this large model space (LNPS) has been developed 
that can reproduce the different phenomena suggested by the available data in 
this mass region~\cite{LNPS}. The LNPS interaction has been obtained by 
adapting a realistic CD-Bonn potential to this model space 
after
many-body perturbation techniques~\cite{MHJ95} and monopole modifications. 
The results for the excitation energy and $B$(E2) values obtained with 
standard effective charges ($e_n=0.5e, e_p=1.5e$) for the Fe isotopes of 
interest are reported in Table~\ref{Table1}.

The excitation energies are very well reproduced by the calculations 
in all cases. 
The calculated $B$(E2)
of $^{64}$Fe coincides with the 
data, 
while overestimated values are obtained for $^{62,66}$Fe.  
The theoretical $B$(E2) values increase linearly with the neutron number but  
the experimental values seem to suggest a slightly flatter behavior. 
  
The analysis of the ground state wave functions shows 
that in $^{62}$Fe, the probability 
of excitations to the $gd$ orbitals is about 60\%, increasing 
to $\sim$~90\% in $^{64}$Fe and to almost 99\% in $^{66}$Fe. 
This is consistent with the development of a new ``island 
of inversion'' at $N$~=~40 in this mass region.
The calculated $B$(E2) values of $^{62,64,66,68}$Fe and $^{60}$Cr
are plotted 
in Fig.~\ref{energies}~(c).
The 
trend is similar to that expected for these isotopes
from the VPS scheme, suggesting that a region of enhanced collectivity
extends to neutron-rich Fe and Cr isotopes beyond $N$~=~40.



In summary, we measured the lifetimes of the $2^{+}_{1}$ states of $^{62,64,66}$Fe,
quantifying for the first time an enhanced collectivity of the Fe isotopes at $N$~=~40.
The observed trends in $E$(2$^{+}$) and $B$(E2) are interpreted by
the valence proton symmetry with respect to the proton number 
$Z$~$\approx$~30, suggesting 
similarities in the role played by active protons or holes 
in the 1$f_{5/2}$ and 1$f_{7/2}$ orbitals, respectively.
The shell model calculations with the new effective LNPS interaction
describe the present $B$(E2) data well. 
The calculations point to the important role
of the neutron intruder configurations where excitations
to the 1$g_{9/2}$ and 2$d_{5/2}$ orbitals across the $N$~=~40
subshell gap are favored due to the shell migration far from stability. 
These
results suggest that, owing to the collaborative interplay
between the valence proton and intruder neutron configurations,
a new deformation region appears in the neutron-rich Fe isotopes at
and beyond $N$~=~40.


\begin{acknowledgments}
The authors thank F. Nowacki, A. Poves and K. Sieja for fruitful discussions. 
This work is supported by 
the US NSF under 
PHY-0606007, 
PHY-0758099, and MRI PHY-0619497,
and also
partly by the DFG (Germany) under 
DE1516/1-1 and 
GSI, F.u.E. 
OK/JOL.
\end{acknowledgments}


\end{document}